# Nasdaq-100 Companies' Hiring Insights: A Topic-based Classification Approach to the Labor Market

Seyed Mohammad Ali Jafari[1[0000-0002-4315-7812]] and Ehsan Chitsaz[2[0000-0002-9656-418X]]

[1] PhD candidate, Technological Entrepreneurship Department, Faculty of Entrepreneurship, University of Tehran, Iran
Sma_jafari@ut.ac.ir
[2]Assistant Professor, Faculty of Entrepreneurship, University of Tehran, Iran,
Chitsaz@ut.ac.ir, corresponding author

**Abstract.** The emergence of new and disruptive technologies makes the economy and labor market more unstable. To overcome this kind of uncertainty and to make the labor market more comprehensible, we must employ labor market intelligence techniques, which are predominantly based on data analysis. Companies use job posting sites to advertise their job vacancies, known as online job vacancies (OJVs). LinkedIn is one of the most utilized websites for matching the supply and demand sides of the labor market; companies post their job vacancies on their job pages, and LinkedIn recommends these jobs to job seekers who are likely to be interested. However, with the vast number of online job vacancies, it becomes challenging to discern overarching trends in the labor market. In this paper, we propose a data mining-based approach for job classification in the modern online labor market. We employed structural topic modeling as our methodology and used the NASDAQ-100 indexed companies' online job vacancies on LinkedIn as the input data. We discover that among all 13 job categories, Marketing, Branding, and Sales; Software Engineering; Hardware Engineering; Industrial Engineering; and Project Management are the most frequently posted job classifications. This study aims to provide a clearer understanding of job market trends, enabling stakeholders to make informed decisions in a rapidly evolving employment landscape.

## 1 Introduction

### 1.1 Background

The labor market's structure is significantly influenced by various factors, including digitalization, often termed digital disruption, which fundamentally changes employment rules and the competencies required of employees (Chinoracký & Čorejová, 2019). Environmental challenges linked to global population growth can be addressed through innovative technologies such as fully electric machines (HEV/EV), which catalyze the emergence of new industries necessitating novel types of employees (López Ibarra, Matallana, Andreu, & Kortabarria, 2019). Moreover, demographic shifts alter labor market demands and firm strategies, influencing the dynamics of employment across sectors (Hopenhayn, Neira, & Singhania, 2022). This complex and unpredictable labor market environment underscores the need for Human Resource (HR) managers, policymakers, and job seekers to gain a nuanced understanding of labor market trends. Recent research has shown that the adoption of machine learning and big data analytics can significantly enhance the analysis of labor market dynamics, providing essential insights for policy makers and HR professionals (Mršić, Jerkovic, & Balkovic, 2020).

LinkedIn is primarily a social network focused on facilitating the hiring process for both job seekers and employers. Many high-tech companies rely on LinkedIn as a tool for talent acquisition and hiring employees. Although LinkedIn is an effective hiring tool (Karakatsanis et al., 2017), it lacks certain features that could help human resource managers, policymakers, and job seekers understand labor market trends and patterns. For instance, it is difficult to discern which job categories companies are most interested in, or to analyze the prevalence of job categories in comparison to one another.

## 1.2 Motivation

Companies during digital transformation and technological improvement should consider answering this question: “What kind of advanced technologies should be used to improve competitive advantage?” One aspect of businesses that should use new technological tools is human resource management. Human resource managers should use more accurate and real-time tools to understand labor market trends and information (Puhovichova & Jankelova, 2020). LinkedIn and other online job vacancy platforms focus on matching job seekers and employers together but do not provide strategic insights for companies and job seekers to understand their industries' labor market trends. This paper proposes a structural topic modeling approach to analyze job advertisement posts from Nasdaq-100 companies (which are heavily high-tech based) on LinkedIn to classify and show their prevalence. Such analyses represent technological advances and can be of great value to human resource managers, policymakers, and job seekers to understand categories and the prevalence of the modern job market.

Human resource managers face challenges in adapting to rapid shifts in employment conditions and maintaining workforce engagement, compounded by the intricacies of legal and organizational frameworks that emerged particularly during the COVID-19 pandemic. These frameworks include handling changes in employment laws, managing compensation issues, and preserving employee well-being (Yusefi et al., 2022). Policymakers are tasked with developing responsive employment programs to stabilize and grow the economy post-disruption, while job seekers navigate an increasingly complex job market, where agility and upskilling become necessary to align with new industry demands.

This study leverages the power of data mining and structural topic modeling to offer a granular analysis of labor market trends within high-tech industries, particularly those represented within the NASDAQ-100. By classifying job postings on LinkedIn, this research not only enhances understanding of the labor market but also aids stakeholders in making informed decisions about career pathways, HR strategies, and policy formulations. This approach is expected to contribute significantly to the literature on labor market intelligence by providing a detailed view of how high-tech companies use digital platforms for recruitment (Grishin, Shaykhutdinov, Gainullin, & Sadykova, 2022).

# 2 Related Works

## 2.1 Overview of Current Literature

Due to the rapid growth of online job vacancy advertisements and their increasing credibility among young talented individuals, we could implement AI methodologies such as Machine Learning (ML) and Natural Language Processing (NLP) to explore and discover trends in the labor market. This kind of analysis, named Labor Market Intelligence (LMI), offers a faster and more expansive capability for analyzing trends compared to traditional surveys (Boselli, Cesarini, Mercorio, & Mezzanzanica, 2018).

In recent research in the field of labor market intelligence, researchers have used various machine learning and natural language processing techniques to classify online job vacancies extracted from different sources, employing standard taxonomies such as the European Skills, Competences, Qualifications and Occupations (ESCO), International Standard Classification of Occupations (ISCO), or Occupational Information Network (ONET). Karakatsanis et al. (2017) utilized a data mining-based approach to identify the most in-demand occupations in the modern job market by employing the Latent Semantic Indexing (LSI) model extracted from the web with occupation description data in the ONET database. Boselli et al. (2018) presented an approach for automatically classifying web job vacancies on a standard taxonomy of occupations via machine learning; they developed a method that could support multi-language input and compared international labor markets with each other. Varelas, Lagios, Ntouroukis, Zervas, Parsons, & Tzimas (2022) presented a hard voting algorithm for the classification of job postings according to the ISCO Occupation Codes.

Another group of researchers focuses on finding the relationship between the demand and supply sides of the labor market. Papoutsoglou, Rigas, Kapitsaki, Angelis, & Wachs (2022) used natural language processing techniques and explored both online job vacancy sites and tech communities to determine each side's focused topics. Aleisa, Beloff, & White (2022) aimed to decrease the selection gap between recruiters and job seekers by using a three-layer AI architecture that utilized clustering algorithms to generate similarity scores between recruiters and job seekers.

Some researchers focus on organizing labor market information as a graph. Giabelli, Malandri, Mercorio, & Mezzanzanica (2020) aimed to achieve two goals through their research: enabling the representation of occupation/skill relevance and similarity over the European Labor Market, and enriching the European standard taxonomy of occupations and skills (ESCO) to better fit the labor market expectations.

### 2.2 Gap in Literature and Proposed Approach

Despite the extensive use of data mining for job classification, there is a notable gap in the application of Structural Topic Modeling (STM) within this context, particularly concerning high-tech industries represented in the NASDAQ-100. Previous research has predominantly focused on broader market analyses without delving into the specific characteristics of job postings in highly specialized sectors (Aleisa, Beloff, & White, 2022; Varelas et al., 2022). This study proposes to fill this gap by applying STM to classify job advertisements from NASDAQ-100 companies on LinkedIn, aiming to uncover nuanced insights into the labor demand in high-tech industries.

### 2.3 Relation to Previous Studies

Our methodology directly builds on the foundations laid by Karakatsanis et al. (2017), who utilized data mining to identify trending occupations, and Boselli et al. (2018), who classified web job vacancies on a standard taxonomy. However, unlike these studies, our approach incorporates additional metadata from LinkedIn, such as company size and sector, to enhance the structural topic models. This adaptation allows for a more detailed analysis of job categories and their prevalence within the high-tech sector, addressing a critical research need highlighted by recent studies. Emerging research in high-tech employment trends reveals the need for continuous monitoring and analysis to better understand sector-specific job dynamics and their broader economic impacts (Chuchkalova & Orekhova, 2021).

### 2.4 Innovations in Topic Modeling for Labor Market Intelligence

Further supporting our approach, recent advancements in topic modeling have shown that incorporating metadata can significantly improve the quality and relevance of the derived topics (Roberts, Stewart, & Tingley, 2019). This methodological enhancement is particularly pertinent in sectors like technology, where the rapid evolution of job roles and skills necessitates sophisticated analytical tools to keep pace with industry changes (Mirza, Mulla, Parekh, Sawant, & Singh, 2015).

## 3 Data Collection and Pre-Processing

### 3.1 Data collection

**NASDAQ-100.** We selected Nasdaq-100 companies to discover labor market trends in flagship high-tech companies in the United States. Firms in the NASDAQ-100 Index are considered the best representation of non-financial securities listed on The NASDAQ Stock Market. The index comprises mostly U.S. firms and is somewhat "tech-heavy," offering opportunities for growth (Rutledge, Karim & Lu, 2016).

Nasdaq-100 indexed companies were selected because their job pages on LinkedIn were very active, thus, by analyzing this index, we could achieve a more meaningful classification. The sectors of Nasdaq-100 companies are demonstrated in Figure 1.

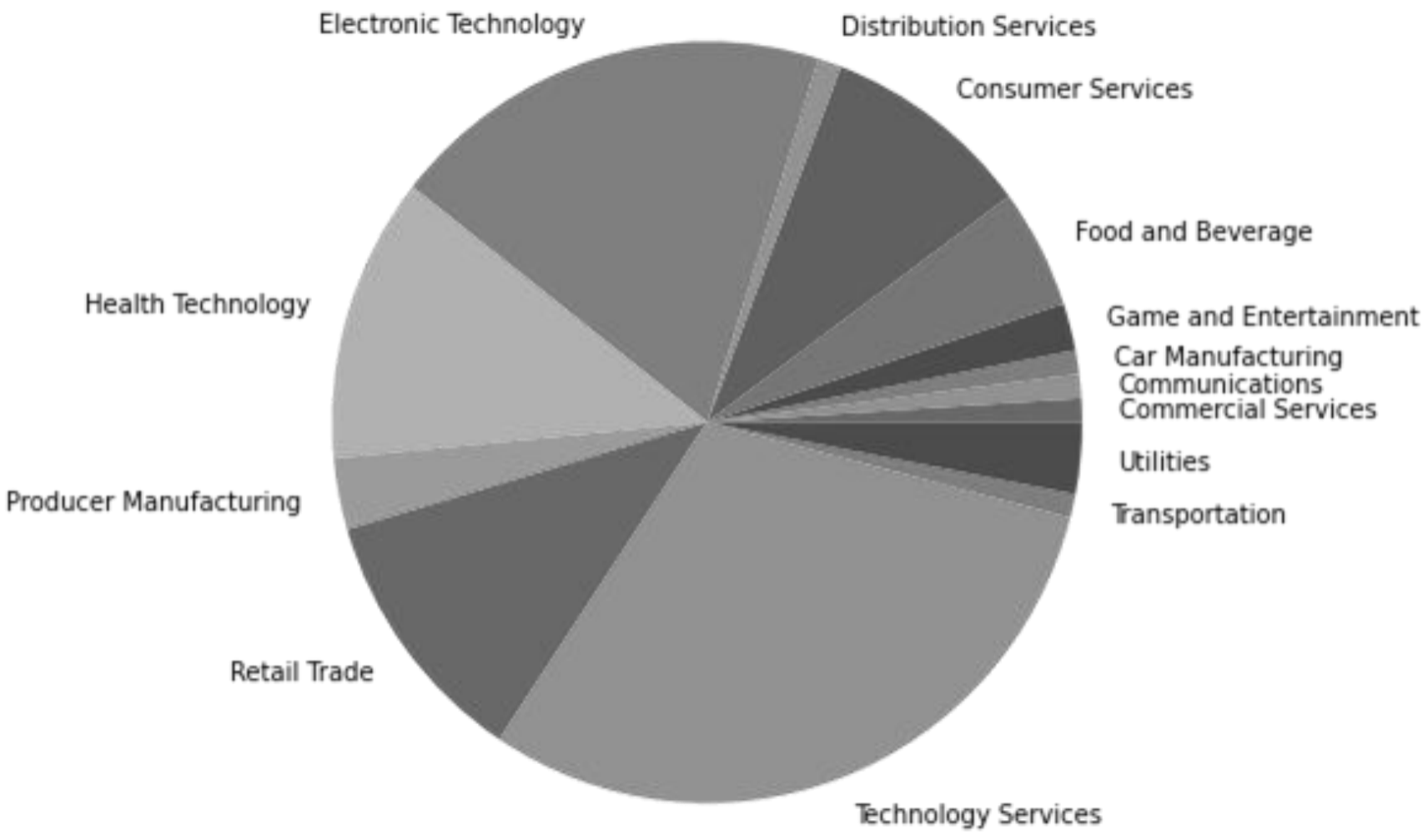

**Fig. 1.** Demonstration of nasdaq-100 companies' sectors

LinkedIn. According to Pejic-Bach, Bertoncel, Meško, Zivko, & Krstić (2020), job advertisements are the most important channel for attracting new employees for the following reasons: LinkedIn is suitable as a data source for research because 1) LinkedIn has become one of the most prominent leaders in publishing job advertisements covering a broad range of organizations, countries, and job types; 2) LinkedIn job advertisements have a semi-structured form, which is suitable for text mining analysis (Pejic-Bach et al., 2020). We wrote a Python program to scrape all jobs that Nasdaq-100 companies released within one month. It is important to mention that LinkedIn does not have historical data and the dates of job posts are not available from LinkedIn. Thus, we faced a situation where LinkedIn did not share the job advertisement date, and the longest timestamp we were able to use for retrieving job advertisements was one month. We chose the November to December timespan because it is an important recruitment season for companies; they complete budgets in October and November and post new jobs in December (Umphress, 2017). We collected 19,961 job postings from Nasdaq-100 companies' LinkedIn job pages within a one-month (November to December 2022) timespan to discover labor market trends and job prevalence among companies. We limited job advertising posts by setting the relevance filter to show the most recent job postings within a month and set the experience level to "None" to display all job positions regardless of job seekers' prior experiences.

The choice of LinkedIn is driven by its extensive use among NASDAQ-100 companies for recruitment, providing a direct view into the qualifications and skills demanded by top technology firms. However, focusing exclusively on these companies and this platform may introduce a selection bias towards high-tech roles and may not reflect the broader industry spectrum. It's important to recognize that the findings from this dataset might not be generalizable to all sectors or job markets.

### 3.2 Pre-processing

To ensure the integrity and quality of the data extracted from LinkedIn, we applied a series of pre-processing steps designed to optimize the data for subsequent analysis. These steps are crucial for enhancing the accuracy of the topic modeling process by reducing noise and focusing on the most relevant information.

**Text Normalization and Cleaning.** All job postings were standardized by converting text to lowercase to eliminate discrepancies caused by case sensitivity. Non-alphanumeric characters, numbers, and short words of fewer than three characters were removed to simplify the text and focus on meaningful content. This normalization process is critical for maintaining consistency across the dataset and for the effective application of natural language processing techniques (Savin et al., 2022).

**Stop Words Removal.** Common English stop words, along with domain-specific jargon that offers minimal informational value, were removed using the Natural Language Toolkit (NLTK). Additionally, we employed a custom list of stop words tailored to our specific dataset, which includes terms frequently found in job postings but that do not contribute significantly to the overall analysis. This step is essential for reducing the dimensionality of the data and enhancing the focus of the topic modeling on significant terms (Sarica & Luo, 2021).

## 4 Proposed Methodology

### 4.1 Topic modeling

**Topic Modeling Overview.** Natural language processing (NLP) has evolved significantly with contributions from both linguistics and computer science, enabling computers to comprehend and analyze human language (Varelas et al., 2022). In the realm of NLP, statistical approaches utilize machine learning and probabilistic models to create robust statistical frameworks for text analysis, which is advantageous over the symbolic approach that relies heavily on predefined rules and can falter with novel or complex inputs (Zhao et al., 2021). This research utilizes statistical NLP techniques, specifically focusing on topic modeling to extract latent topics from job advertisements of NASDAQ-100 companies (Papoutsoglou et al., 2022).

**Spectral Classification vs. Dirichlet Allocation.** Among the various algorithms developed for topic modeling, Latent Dirichlet Allocation (LDA) is well-known and widely used in academic research for its ability to model topics as distributions over words (Mustak, Salminen, Plé & Wirtz, 2021). However, for this study, we also employed the spectral classification technique as implemented in the R package by Roberts, Stewart, & Tingley (2019). Spectral classification, by default, offers advantages in topic clarity and separation by utilizing matrix decomposition techniques that are often more effective in identifying distinct topic boundaries compared to LDA. To empirically validate the effectiveness of spectral classification over LDA, we conducted a comparative analysis using both methods on the same dataset. Our findings indicate that spectral classification provides enhanced accuracy and coherence in topic discovery, supporting its selection for this analysis (Hannigan et al., 2017).

**Challenges and Limitations of STM.** While STM offers several advantages, such as the ability to incorporate metadata which can enrich the model's contextuality, it also presents challenges. These include the need for careful selection of metadata and the risk of overfitting if not properly tuned. Additionally, STM can be computationally intensive, particularly with large datasets, which requires careful consideration of computational resources (Savin et al., 2022).

**Selection Criteria for Topic Models.** The process of determining the optimal number of topics, denoted as 'K', is critical in topic modeling. For this study, we experimented with various models ranging from 10 to 50 topics, evaluating each model's semantic coherence and exclusivity—a measure of how well topics are differentiated from one another. The analysis revealed that models with 10 to 20 topics provided the best balance of coherence and distinctiveness. After further refinement, we found that a model with 13 topics struck the optimal balance, exhibiting the highest levels of semantic coherence and exclusivity. This selection was corroborated by testing additional models in smaller increments, confirming that 13 topics provided the most insightful and interpretable results (Papoutsoglou et al., 2022).

### 4.2 Validation

Validating a topic model derived from natural language data involves tackling inherent complexities due to the high dimensionality and nuanced features embedded within the data (Chan, Rao, Huang, & Canny, 2019). In our study, we employed t-distributed Stochastic Neighbor Embedding (t-SNE) to facilitate this process. t-SNE is renowned for its effectiveness in reducing high-dimensional data to lower-dimensional spaces, making it possible to visually assess the coherence and accuracy of clustering (Cao & Wang, 2017).

**Rationale and Advantages of Using t-SNE.** We chose t-SNE over other dimensionality reduction techniques like PCA or MDS due to its exceptional ability to maintain local data structures, which is crucial for accurately reflecting the thematic closeness of job postings. This method allows for a visual inspection of how well the topic model groups similar content, which is indispensable for our analysis (van der Maaten & Hinton, 2008).

**Performance Metrics and Enhanced t-SNE Usage Discussion**. To quantitatively validate the topic model, we utilized two intrinsic metrics: coherence and exclusivity. Coherence evaluates how semantically related the top words within each topic are, providing an indication of how interpretable these topics are to human analysts. Exclusivity measures the uniqueness of the top words to their respective topics, enhancing the model's ability to distinctly categorize content (Savin et al., 2022).

These metrics were complemented by the t-SNE visualization, which plots each job posting as a point in a two-dimensional space. The effective clustering of these points according to their topics not only visualizes but also substantiates the model's accuracy. We specifically looked for tightly-knit clusters within the visualization as these indicate that the model has successfully grouped postings with similar topical themes. The proximity of postings within the same cluster in this reduced space supports the validity of our model, affirming that postings predicted to be similar are indeed closely related (Mustak et al., 2021).

**Visualization and Interpretation**. In Figure 2, we visualize the clustering outcome using t-SNE, where each of the 19,961 points represents an individual job posting categorized into one of 13 distinct topics. This visual representation helps validate the clustering efficacy of our topic model by clearly showing grouped postings within distinct areas of the plot.

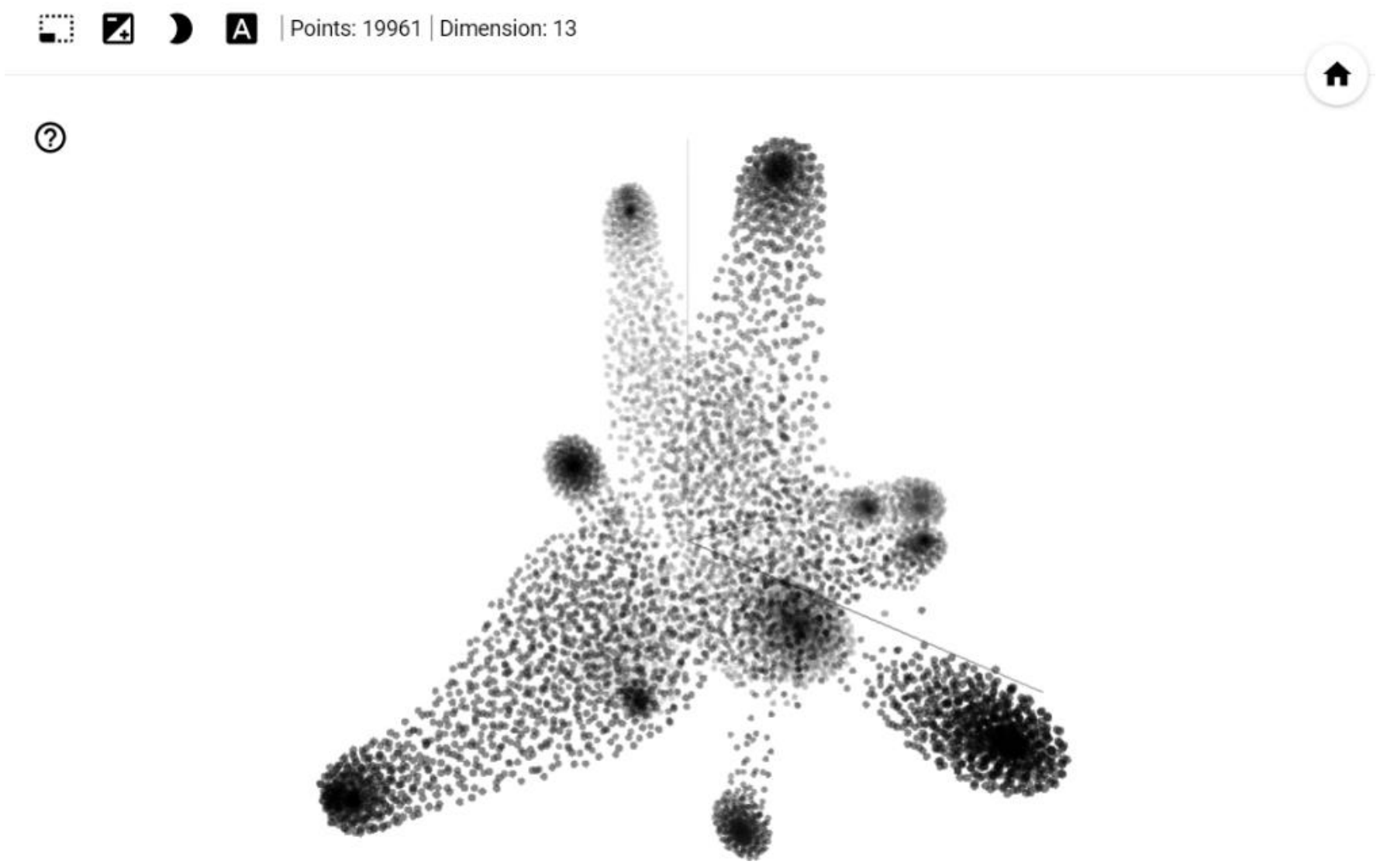

**Fig. 2.** T-distributed Stochastic Neighbor Embedding (t-SNE) of the modeled topics

## 5 Result and Discussion

Table 1 demonstrates 13 topics achieved with help of the Structural Topic Modeling (STM) R package, it also shows the prevalence of each topic among all 19961 job descriptions and the most discriminating (frequent and exclusive) words within each topic. Figure 3 shows the topics' prevalence in comparison with each. an easier way for understanding topics' subjects is to use word clouds with word clouds we could visually see the distribution of words among a collection of job descriptions. Figure 4 represents word clouds of each topic to help us recognize the subject of each topic.

**Table 1.** Topics identified for job advertisement descriptions

| No | Topic label | Words | Prevalence |
|---|---|---|---|
| 1 | Distribution and Transportation | Customer, vehicle, service, inventory, support, distribution | 5.0 |
| 2 | Trash (non english words) | Des, experiencia, het, vous | 1.8 |
| 3 | Marketing, Branding, and Sales | Business, management, marketing, strategy, customer | 16.9 |
| 4 | Industrial Engineering and Project Management | Management, engineering, project, energy, technical, power | 10.4 |
| 5 | Data Analytics Solutions for Service Sector | Data, business, details, analysis, analytics | 2.7 |
| 6 | General Internal Business Management | Management, medical, business, support, process | 5.6 |
| 7 | Medical Development and Manufacturing | Clinical, development, manufacturing, medical, support, data | 7.4 |
| 8 | Operational Management | Sales, store, manager, customer, clients | 9.2 |
| 9 | Hospitality and Tourism Services | Guests, service, international, property, guest, food, hotel | 4.8 |
| 10 | Data Analytics Solutions for Product and Sales | Data, business, financial, finance, analytics, insights | 8.0 |
| 11 | Game and Content Development | Design, games, game, content, entertainment | 4.5 |
| 12 | Software Engineering | Software, technical, engineering, development, design | 13.1 |
| 13 | Hardware Engineering | Design, engineering, development, test, verification, semiconductor, hardware | 10.6 |

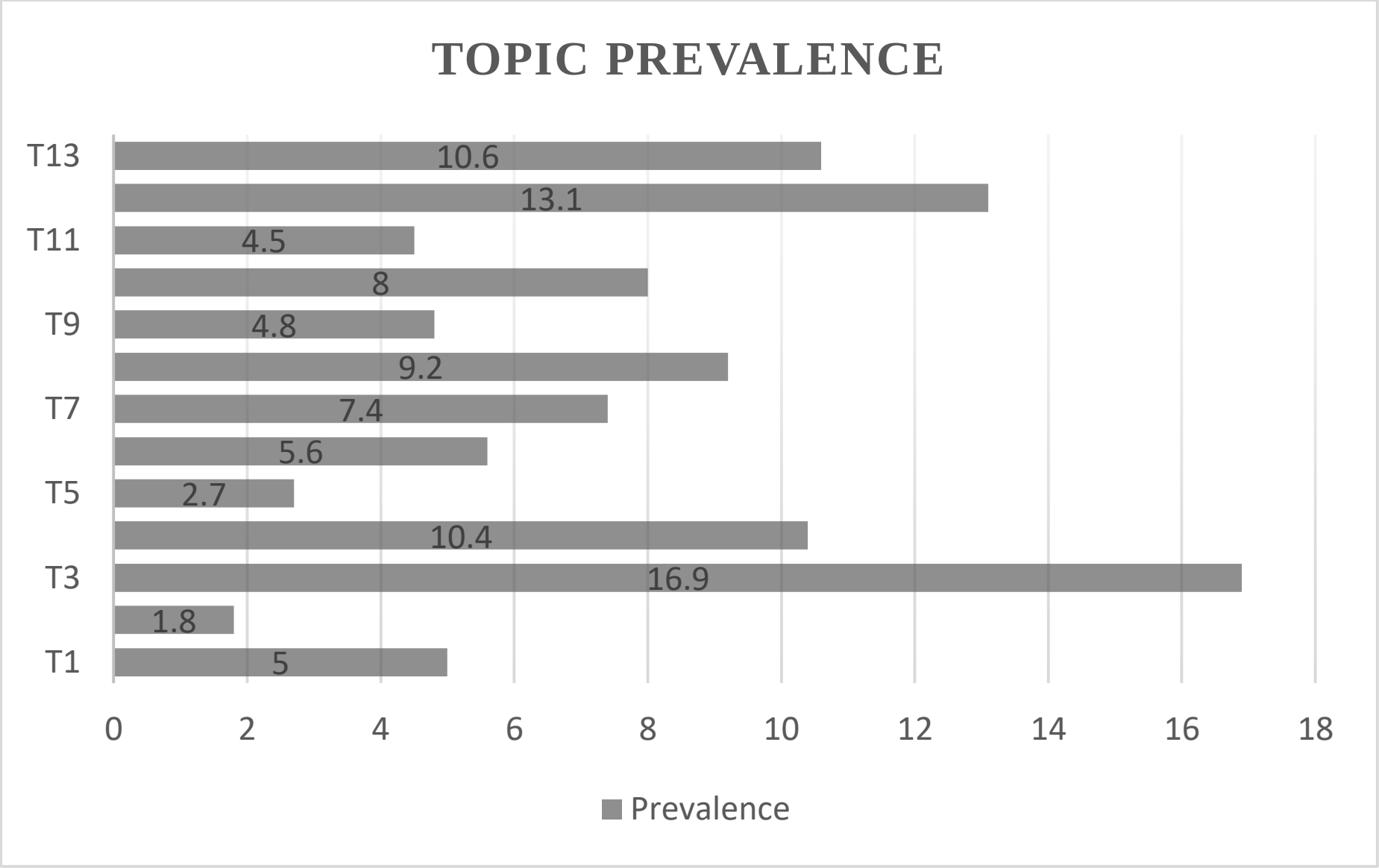

**Fig. 3.** 13 topics prevalences in whole job advertisements

| T1: Distribution and Transportation | T2: Trash (non english words) | T3: Marketing, Branding, and Sales |
|---|---|---|
| 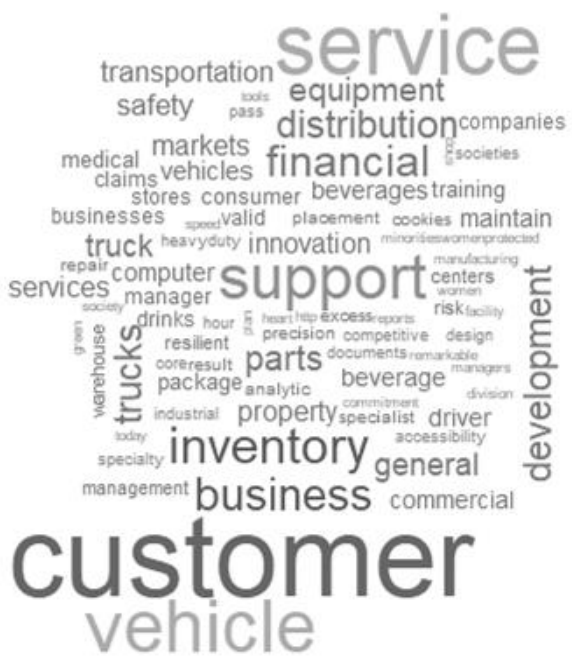 | 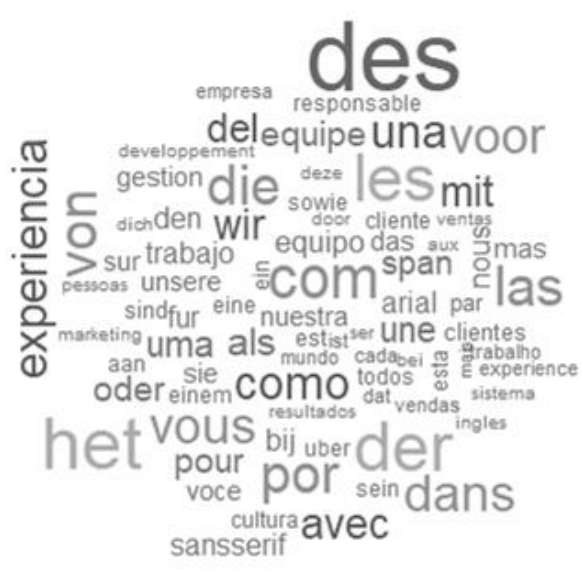 | 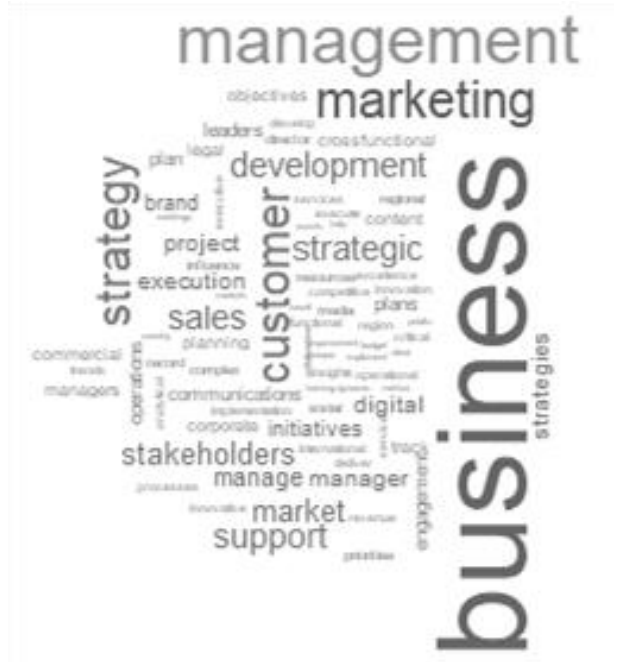 |
| T4: Industrial Engineering and Project Management | T5: Data Analytics Solutions for Service Sector | T6: General Internal Business Management |

<table>
<tr>
<td>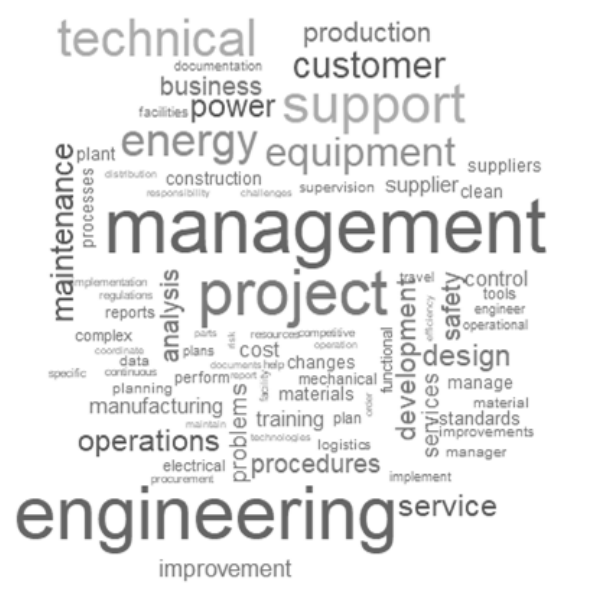
</td>
<td>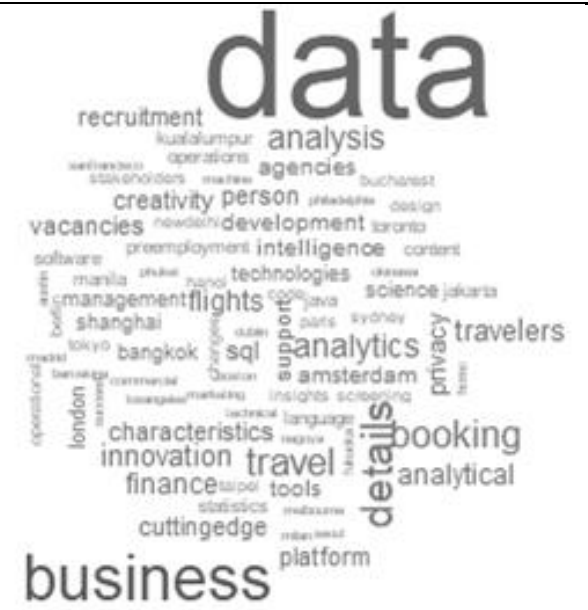
</td>
<td>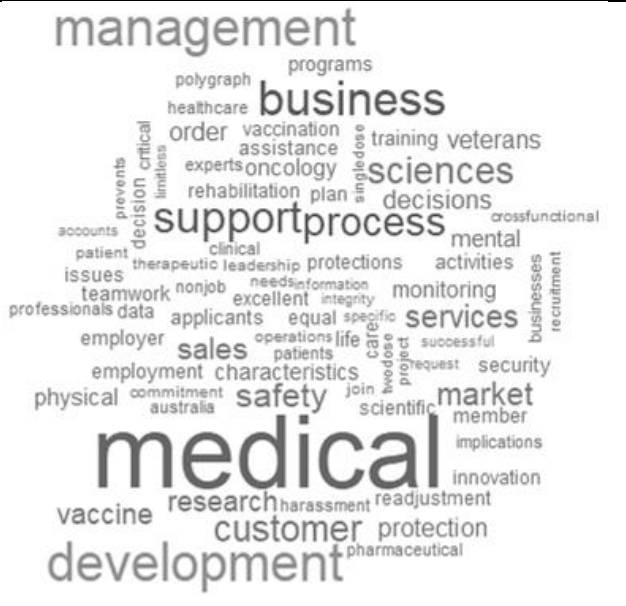
</td>
</tr>
<tr>
<td>T7: Medical Development Manufacturing<br>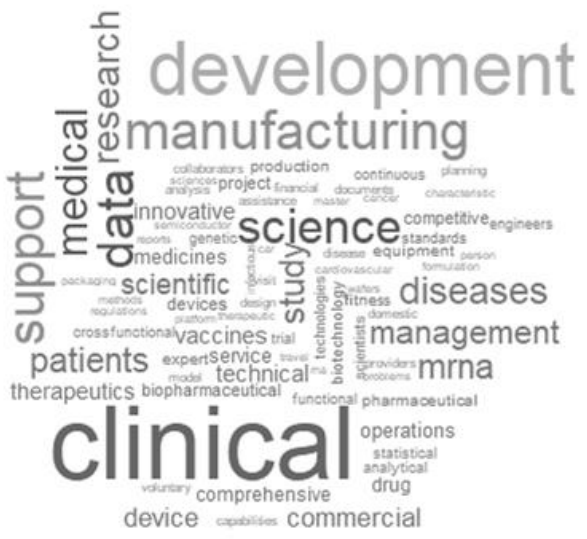
</td>
<td>T8: Operational Management<br>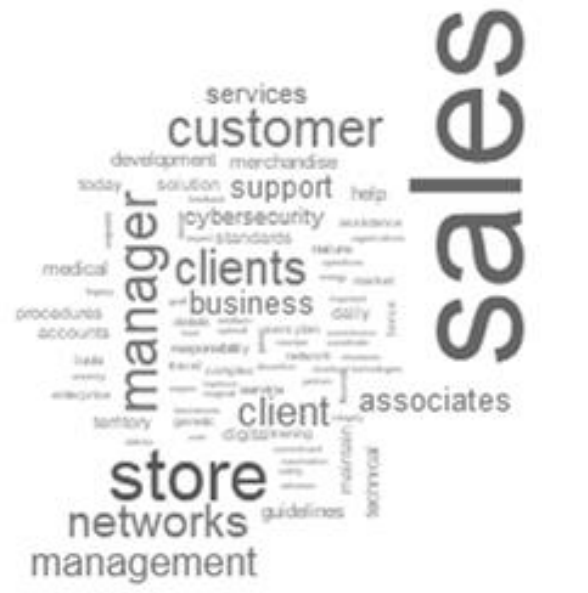
</td>
<td>T9: Hospitality and Tourism Services<br>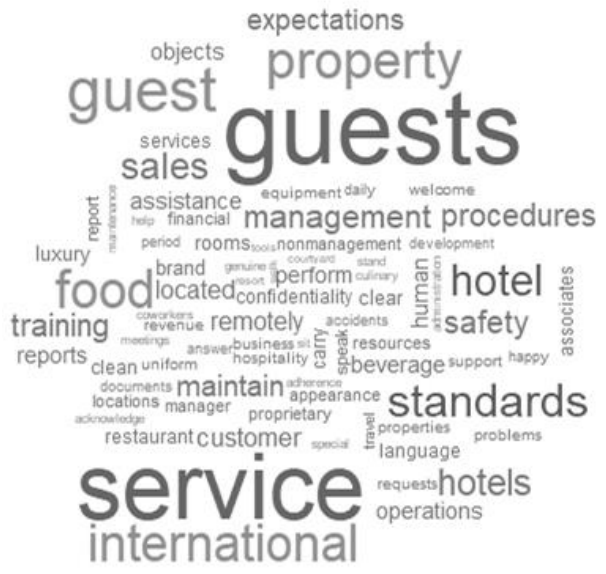
</td>
</tr>
<tr>
<td>T10: Data Analytics Solutions for Product and Sales<br>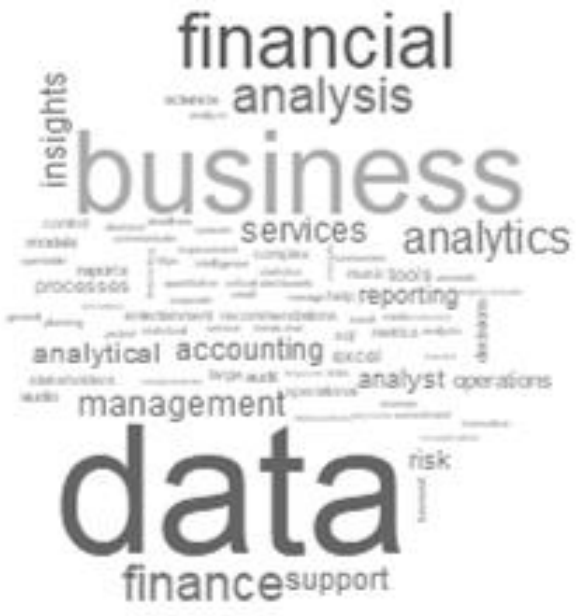
</td>
<td>T11: Game and Content Development<br>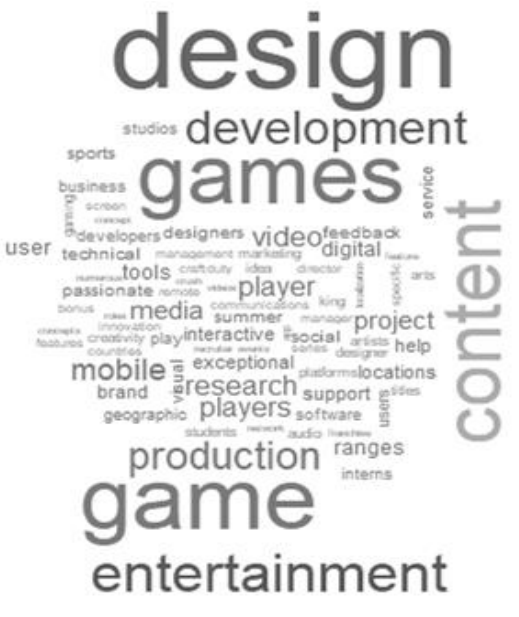
</td>
<td>T12: Software Engineering<br>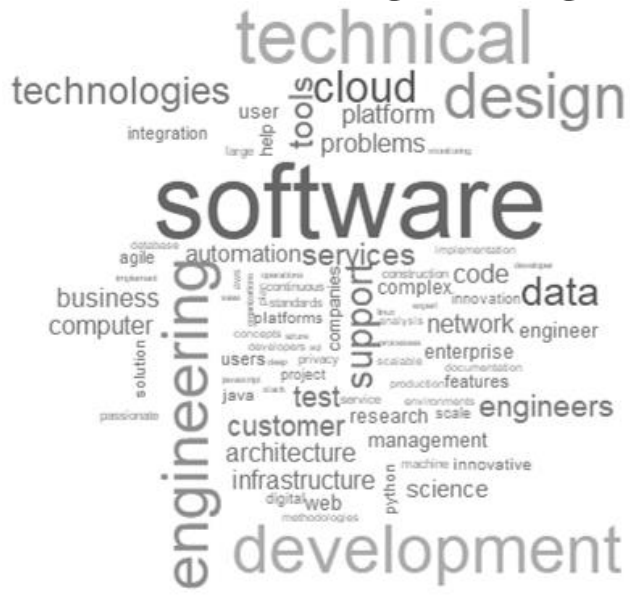
</td>
</tr>
<tr>
<td colspan="3">T13: Hardware Engineering</td>
</tr>
</table>

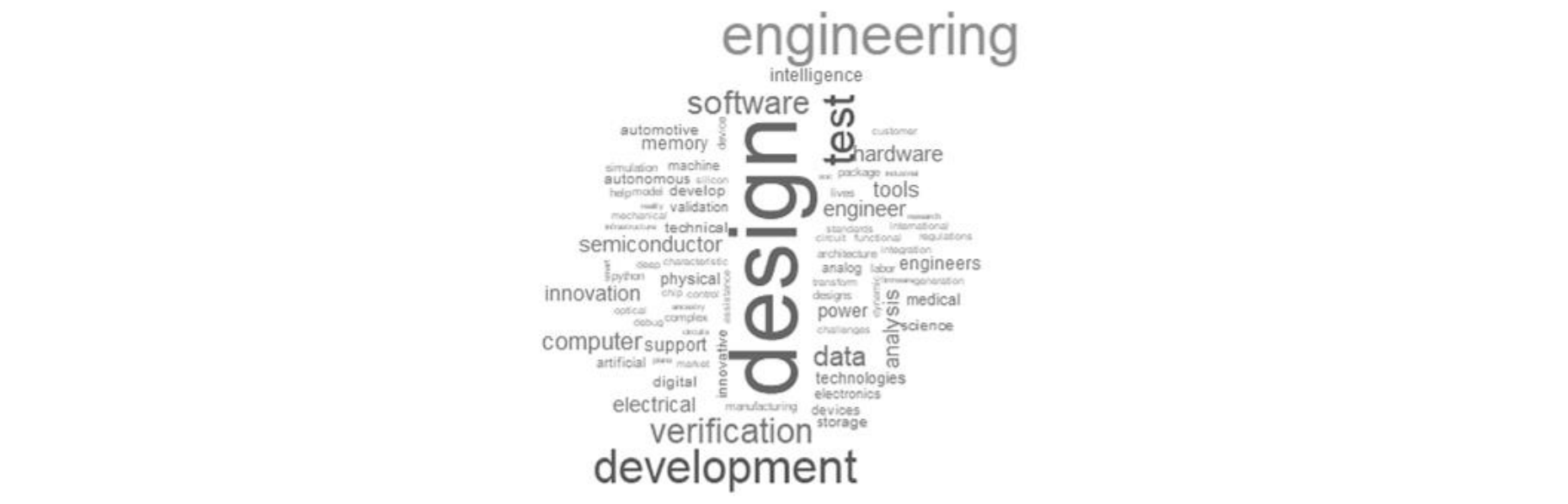

**Fig. 4.** Word clouds of 13 topics were generated based on full descriptions of nasdaq-100 companies' LinkedIn job advertisements. Note: The font size corresponds to the probability (weight) of the respective word given.

We now discuss each topic individually. we filtered job advertisements (documents) with a threshold of 50 percent of relevance to each topic. as mentioned before each document has a probability distribution of all topics with different percentages for example an online job vacancy document (job vacancy) 3 has 50% T1, 25% T2, 10%, T3, …, 0.09% T13 topic relevancy. for example, an online job vacancy with the "Sales Associate" title has 80.93 percent relevance to the Marketing, Branding, and Sales category. after gathering all jobs with more than 50 percent relevance to each topic we picked examples of each topic related job posts and wrote along topic names.

**T1- Distribution and Transportation.** This job category contains jobs related to Distribution and Transportation. An example online job vacancy description that Copart posted with Inventory Specialist title is as follows: "the Inventory Specialist will be responsible for facilitating the Copart experience by offering solutions to meet customer's needs. Monitor, maintain and organize the receiving area - Operate camera and utilize a handheld inventory device to process incoming vehicles - Determine operational capability of motor vehicles - Complete vehicle inspection inventories (TLEs) on required vehicles - Maintain inventory of all materials used… "

**T2- Trash (non english words).** Some multinational companies posted jobs in local languages, for example, an Align technology job's advertisement had been written in Portuguese because local people were targeted. (1265) "O Gerente de Território é responsável pelo desenvolvimento e gestão de vendas de um determinado território, buscando proativamente avaliar as necessidades e o potencial de crescimento de uma base de clientes…"

**T3- Marketing, Branding, And Sales.** This job category contains jobs related to Marketing, Branding, And Sales. An example online job vacancy description that Mondelēz International posted with Consumer Experience Manager, Chocolates, India & Bangladesh title is as follows: "One of the strategic pillars of this is constantly improving our media ROI by maximizing incremental net revenue from media. Main Responsibilities: Develop robust media strategies for the Chocolates Category in line with global media principles and overall company objectives…"

**T4- Industrial Engineering and Project Management.** This job category contains jobs related to Industrial Engineering and Project Management. An example online job vacancy description that Constellation posted with Manager Engineering (Kennett Square) title is as follows: "Primary Purpose of Position: Provide management of the engineering function of accountability. Primary Duties and Accountabilities: Provide management of the engineering function with respect to station needs and regulatory requirements. (35%) Manage the performance and development of assigned engineering personnel relative to site and corporate objectives and provide focus on the attainment of high-quality engineering results. (30%)…"

**T5- Data Analytics Solutions for Service Sector.** This job category contains jobs related to Data Analytics Solutions for Service Sector. An example online job vacancy description that Agoda posted with Senior Business Analyst (Bangkok Based, relocation provided) title is as follows: "In this Role, you'll get to Search: Experiment with text ads, bidding, and campaign structures on Google, Bing, Baidu, Naver, and other search engines. Adapt to new product features and roll out changes from successful tests Display: Test, analyze, and optimize campaigns on Facebook, Twitter, Instagram, and others Modeling: Analyze the vast amounts of data generated by experiments, develop models we can use for optimization, and build dashboards for account managers…"

**T6- General Internal Business Management.** This job category contains jobs related to General Internal Business Management. An example online job vacancy description that Cintas posted with General Manager title is as follows: "Cintas is seeking a General Manager to manage operations while implementing appropriate Cintas policies and procedures. Responsibilities include hiring, coaching, developing and leading a team of service and administrative professionals…"

**T7- Medical Development and Manufacturing.** This job category contains jobs related to Medical Development and Manufacturing. An example online job vacancy description that Moderna posted with Director, Manufacturing Science and Technology title is as follows: "This position is part of Moderna's Manufacturing Science and Technology (MST) team responsible for ensuring robust production of Drug Substance (DS) using our mRNA manufacturing platform at our Norwood site…"

**T8- Operational Management.** This job category contains jobs related to Operational Management. An example online job vacancy description that Dollar Tree Stores posted with operational assistant manager title is as follows: "Responsible for assisting with all operational tasks within the store as delegated and assigned by the Store Manager with main focus on the front-end and sales floor operations…"

**T9- Hospitality and Tourism Services.** This job category contains jobs related to Hospitality and Tourism Services. An example online job vacancy description that The Ritz-Carlton Hotel (marriott-international) posted with Guest Relations Agent - The Ritz-Carlton Maldives, Fari Islands title is as follows: "Answer, record, and process all guest requests, questions, or concerns via telephone, email, chat, and mobile communication devices…"

**T10- Data Analytics Solutions for Product and Sales.** This job category contains jobs related to Data Analytics Solutions for Product and Sales. An example online job vacancy description that Pandora (siriusxm) posted with Business Analyst title is as follows: "SiriusXM Media is looking for a Business Analyst to play a critical role in evaluating our ad product technology…"

**T11- Game and Content Development.** This job category contains jobs related to Game and Content Development. An example online job vacancy description that king (activision-blizzard) posted with UI Designer (New Games) title is as follows: "We are looking for a dedicated and multi-skilled UI Designer with a passion for games, graphic design and excellent user experiences. In this role, you will be a core component of UI design in our New Games business working on one of our exciting new projects…"

**T12- Software Engineering.** This job category contains jobs related to Software Engineering. An example online job vacancy description that Datadog posted with Software Engineer - Live Streaming Storage (C++) title is as follows: "We are looking for an Engineer who is interested in distributed data stores, scalability, availability and cares about giving our customers the best platform for them to explore their data. You will: Build distributed, high-throughput, event database - Do it in modern C++ - Own meaningful parts of our service, have an impact, grow with the company…"

**T13- Hardware Engineering.** This job category contains jobs related to Hardware Engineering. An example online job vacancy description that Micron Technology posted with JR36580 Chemical Systems Engineer - Facilities title is as follows: "Micron is looking for a Chemical System Engineer to join our team. In this role, you will lead the design and optimization of chemical distribution systems at the Boise, Idaho campus…"

## 6 Conclusion

The purpose of this study is to gain insights into companies indexed in the Nasdaq-100, which is heavily tech-based (Rutledge et al., 2016). These insights are based on the descriptions of job advertisements from Nasdaq-100 companies. We collected 19,961 Online Job Vacancies (OJVs) from job advertisement descriptions of these companies between November and December 2022.

We selected this timeframe for two reasons: 1) November to December is a crucial recruitment season for companies. They typically complete budgets in October and November and post new jobs in December. Companies with strategic hiring plans tend to seek talent in November and December and expect to make hires in January and February (Umphress, 2017). 2) The maximum time span provided by LinkedIn is one month.

Using structural topic modeling, we have identified 13 job categories from all the Nasdaq-100 companies' job advertisements during the recruitment season.

The insights we provide could be valuable for job seekers, human resource managers, and policymakers. Job seekers can benefit from these insights by better understanding their career paths and the trends within their companies of interest in the labor market. This knowledge could help them enhance their skills to align with labor market preferences. Human resource managers can benefit from these insights by understanding the labor market situation and the strategies of their competitors and market leaders. It is worth mentioning that the companies we studied are from the Nasdaq-100, which are highly valuable in terms of market capitalization within their industries; knowledge of their strategies and actions can provide valuable information about the industries in which they operate. Policymakers could use our findings to make the education system more productive and market-oriented. For example, they could invest more in fields such as marketing, business management, and engineering rather than other majors to accelerate the alignment between the demand and supply segments of the labor market. Job seekers could also align their knowledge with market needs to derive greater benefit from it.

Our study could be enriched in many ways in future studies, 1- Integrating Diverse Data Sources and AI Technologies: Future research could significantly benefit from incorporating a wider array of data sources. While LinkedIn offers a rich dataset, platforms like Stack Exchange and GitHub provide unique insights into the tech community's discussions and collaborations, which are invaluable for understanding the supply side of the labor market. Moreover, adopting Explainable AI (XAI) techniques could enhance the transparency and reliability of AI applications in LMI, making the findings more actionable and trustworthy (F. Colace et al., 2019). 2- Predictive Modeling and Real-Time Analysis: Building predictive models that can analyze trends over longer periods will provide deeper insights into labor market dynamics. Utilizing machine learning frameworks to analyze data collected from various job portals can help forecast future labor demands, offering a proactive tool for job seekers and policymakers (N. Alsayed & W. Awad, 2023). Such models could also integrate real-time economic indicators to dynamically adjust predictions based on current market conditions. 3- Collaborative Frameworks for Multi-Sector

Analysis: Lastly, establishing collaborative frameworks involving academia, industry, and government can enhance the effectiveness of LMI systems. These collaborations can facilitate the sharing of data and insights, leading to more comprehensive labor market analyses that span multiple sectors and geographies. Such efforts will be crucial in developing adaptive educational programs and workforce development initiatives that are aligned with future job market requirements.

As for limitations, our study is dependent on the quality and availability of online job vacancies. Topic modeling classifications are very sensitive to input data, and the quality and precision of classification improve as the quality and specificity of job posts increase. Ideally, online job vacancies should contain detailed job descriptions; however, many are written in a templated format that includes generic information about companies, workplace atmosphere, and employee expectations, which are not very useful for job classification. To overcome this limitation, researchers should work on developing cleaning algorithms that eliminate less pertinent information in job descriptions.